\documentstyle[pre,aps]{revtex}

\title{Fluid particle dynamics: a synthesis of dissipative
particle dynamics and smoothed particle dynamics}
\author{Pep Espa\~{n}ol\footnote{e-mail: pep@fisfun.uned.es}}
\address{Departamento de F\'{\i}sica Fundamental, UNED,
Apartado 60141, 28080 Madrid, Spain}
\date{\today}
\begin{document}

\pacs{
\Pacs{47}{11$+$j}{Computational methods in fluid mechanics}
\Pacs{05}{70 Ln}{Non-equilibrium thermodynamics, irreversible processes}
      }

\maketitle

\begin{abstract}
We present a generalization of dissipative particle dynamics
that includes shear forces between particles. The new
algorithm has the same structure as the (isothermal) smoothed particle
dynamics algorithm, except that it conserves angular momentum and
includes thermal fluctuations consistently with the principles
of equilibrium statistical mechanics. This clarifies the connection
of dissipative particle dynamics with numerical resolution algorithms
of the macroscopic Navier-Stokes equations.
\end{abstract}

Dissipative particle dynamics (DPD) is an off-lattice simulation
technique that has been introduced by Hoogerbrugge and Koelman in
order to address hydrodynamic problems in complex fluids
\cite{hoo92,koe93}.  The technique has received substantial
theoretical support \cite{esp95,hydro} and it has been
successfully applied in a large variety of systems including flow in
porous media \cite{koe93}, colloidal suspension \cite{koe93,boe97},
dilute polymeric suspensions \cite{sch95}, and immiscible binary fluids
\cite{cov96b}.

The idea behind DPD is to simulate a Newtonian fluid (as, for example,
the Newtonian solvent in colloidal or polymeric suspensions) in terms
of mesoscopic ``lumps'' or ``droplets'' of fluid, named dissipative
particles \cite{hoo92,esp97b}. It is postulated that these dissipative
particles interact with each other with a pair-wise conservative
potential, with dissipative forces that depend on the relative {\em
approaching} velocity of the particles, and with random forces that
satisfy a fluctuation-dissipation theorem \cite{esp95}.  Newton's
third law is satisfied and the total momentum of the system is
conserved, although energy is not \cite{hydro}. This implies that
there are local conservation equations for mass and momentum, i.e. the
system behaves hydrodynamically at long times and distances
\cite{hydro}. The advantage of DPD over conventional molecular
dynamics relies on the fact that DPD is a coarse-grained technique
which captures the gross features of mesoscopic portions of fluid.
The microscopic details, which are computationally expensive and not
even interesting, are averaged out in DPD.

The spirit of DPD turns out to be quite similar to that of smoothed
particle dynamics SPD \cite{luc77}. This technique was originally
intended to simulate astrophysical non-viscous flows and has been
recently been applied in a variety of non-viscous
\cite{hoo96b},\cite{sim96} and viscous problems
\cite{tak94},\cite{hoo96}. SPD consists on a discretization of the
Navier-Stokes equations in a Lagrangian moving grid with the aid of a
weight function. In this way, the nodes of the grid can be identified
as ``smooth particles'' interacting through prescribed laws of force, thus
allowing to solve Navier-Stokes equations with molecular dynamics
codes. The Lagrangian nature of the technique makes it very
appropriate to study flows in complex geometries (like those appearing
in colloidal suspensions) because there is no need of costly
recalculations of the mesh as the boundary conditions
evolve. Actually, the very dynamics takes care of it. Unfortunately,
there is at present no implementation of the thermal fluctuations
present in fluids at mesoscopic scales and which are the responsible
for the Brownian motion of small suspended objects. It is not clear
that the fluctuations that appear as a consequence of the discrete
nature of the technique are compatible with the principles of
statistical mechanics. In particular, that they obey a
fluctuation-dissipation theorem. In other words, there is yet no
implementation of SPD for fluctuating hydrodynamics \cite{lan59}.

In this letter we present a model of fluid particles that interact
with dissipative forces that, besides the dissipative force of the
original DPD algorithm, include {\em shear forces} between fluid
particles.  We regard this as a more realistic model in view of the
fact that the proposed algorithm coincides in structure with the SPD
algorithm. Therefore, it is expected that an even more reduced number
of particles already reproduce the hydrodynamic behavior of the
system.  We also formulate the random forces between fluid particles
in such a way that the distribution function of velocities is
Gaussian, as predicted by equilibrium statistical mechanics. In this
way, this work represents a generalization of SPD that includes
thermal fluctuations. This opens up the possibility of applying a
technique closely related to SPD to the study of complex fluids.

A shear force between particles $i,j$ is proportional to the relative
velocities ${\bf v}_{ij}$ of both particles, whereas the dissipative
force in the original DPD algorithm is proportional to the approaching
velocity $({\bf e}_{ij}\!\cdot\!{\bf v}_{ij}){\bf e}_{ij}$, where
${\bf e}_{ij}$ is the unit vector in the joining line of both
particles. The initial motivation for modifying the original algorithm
of DPD by introducing shear forces was the identification of an
elementary motion between dissipative particles that produces no force
in that algorithm. If a dissipative particle is orbiting in a
circumference around a reference particle, it will not exert any force
on this particle. Nevertheless, on simple physical grounds one expect
that the motion of the dissipative particle must drag in some way the
reference particle. This is taken into account through the shear
forces in the fluid particle model presented in this letter.  We note,
however that this relative motion might produce a drag even in the
original DPD algorithm if many DPD particles are involved
simultaneously. The same is true for a purely conservative molecular
dynamics simulation. The point is, of course, that the effect is
already captured with a much smaller number of particles in the fluid
particle model.

The shear forces are not central and therefore angular momentum is not
conserved. We restore angular momentum conservation by including in
the description a spin variable.  If one thinks of the fluid particles
as mesoscopic portions of fluid, this spin variable has a sounded
physical interpretation: it describes the angular momentum of the
molecules that constitute the fluid particle with respect to the
center of mass of the fluid particle.

\section{The fluid particle model}
The fluid particle model is defined by $N$ identical particles of mass
$m$ and moment of inertia $I$. 
The state of the system is
characterized by the positions ${\bf r}_i$, velocities ${\bf v}_i$,
and angular velocities ${\bf \omega}_i$ of each particle.  We do not
include here an internal energy variable and the resulting algorithm,
like DPD, will not conserve energy locally. This may be a minor
problem when one is interested only in rheological properties.

The equations of motion of the system are given by

\begin{eqnarray}
\dot{{\bf r}}_i&=&{\bf v}_i
\nonumber\\
\dot{{\bf v}}_i&=&\frac{1}{m}\sum_{j\neq i}{\bf F}_{ij}
\nonumber\\
\dot{{\bf \omega}}_i&=&\frac{1}{I}\sum_{j\neq i}{\bf N}_{ij}
\label{eqmotion}
\end{eqnarray}
where ${\bf F}_{ij},{\bf N}_{ij}$ are the force and torque that particle
$j$ exerts on particle $i$.  We require that the forces satisfy
Newton's third law, ${\bf F}_{ij}=-{\bf F}_{ji}$, in such a way that
the total linear momentum ${\bf P}=\sum_i m {\bf v}_i$ is a dynamical
invariant, $\dot{\bf P}=0$. In addition, we assume that the torques in
(\ref{eqmotion}) are given by $ {\bf N}_{ij}=-{\bf r}_{ij}\times{\bf
F}_{ij}/2$ and one checks immediately that the total angular momentum
${\bf J}=\sum_i({\bf r}_i\times{\bf p}_i + I\omega_i)$ is conserved
exactly, $\dot{{\bf J}}=0$.

We  model the force ${\bf F}_{ij}$ between fluid particles
according to ${\bf F}_{ij}={\bf F}^C_{ij}+{\bf F}^T_{ij}+{\bf
F}^R_{ij}+{\tilde{\bf F}}_{ij}$ where the different contributions are
given by

\begin{eqnarray}
{\bf F}^C_{ij} &=& -V'(r_{ij}){\bf e}_{ij}
\nonumber \\
{\bf F}^T_{ij} &=& -\gamma m {\bf M}^T({\bf r}_{ij})\!\cdot\!{\bf v}_{ij}
\nonumber \\
{\bf F}^R_{ij} &=& -\gamma m {\bf M}^R({\bf r}_{ij}) \!\cdot\!
\left(\frac{{\bf r}_{ij}}{2}\times[{\bf \omega}_i+{\bf \omega}_j] \right) 
\nonumber\\
{\tilde{\bf F}}_{ij} dt &=& \sigma m
\left({\tilde A}(r_{ij}){\overline{d{\bf W}}}^S_{ij}
+{\tilde B}(r_{ij})\frac{1}{D}{\rm tr}[d{\bf W}_{ij}]{\bf 1}
+{\tilde C}(r_{ij}) d{\bf W}^A_{ij}\right)
\!\cdot\!{\bf e}_{ij}
\label{sum3}
\end{eqnarray}
The first contribution ${\bf F}^C_{ij}$ is a repulsive conservative
force derived from a soft potential $V(r)$. If only this force is
present we have the version of SPD for non-viscous flows studied
extensively in Refs. \cite{hoo96b} ({\em i.e.} a MD simulation). The
second contribution ${\bf F}^T_{ij}$ is a friction force that depends
on the relative translational velocities ${\bf v}_{ij}={\bf v}_i-{\bf
v}_j$. The dimensionless matrix ${\bf M}^T({\bf r}_{ij})$ is the most
general matrix that can be constructed out of the vector ${\bf
r}_{ij}={\bf r}_i-{\bf r}_j$, this is ${\bf M}^T({\bf r}_{ij}) \equiv
A(r_{ij}){\bf 1}+B(r_{ij}){\bf e}_{ij}{\bf e}_{ij}$ where ${\bf 1}$ is
the unit matrix, ${\bf e}_{ij}={\bf r}_{ij}/r_{ij}$ is the unit vector
joining the particles, $r_{ij}=|{\bf r}_{ij}|$ and the functions
$A(r)$ and $B(r)$ provide the range of the force. The friction
coefficient $\gamma$ has been introduced as an overall factor for
convenience and has dimensions of inverse of time.  ${\bf F}^T_{ij}$
is the sum of a shear force $-\gamma m A(r) {\bf v}_{ij}$ and a
central force $-\gamma m B(r) ({\bf e}_{ij}\!\cdot\!{\bf v}_{ij}){\bf
e}_{ij} $.  The rotational contribution ${\bf F}^R_{ij}$ in
(\ref{sum3}) is given also in terms of a dimensionless matrix ${\bf
M}^R={\bf M}^R({\bf r}_{ij}) =C(r_{ij}){\bf 1} +D(r_{ij}){\bf
e}_{ij}{\bf e}_{ij} $ where $C(r),D(r)$ are scalar functions (even
though the $D(r)$ contribution to ${\bf F}^R_{ij}$ is zero we maintain
this term to keep the analogy between ${\bf M}^R$ and ${\bf
M}^T$). Note that if particles $i$ and $j$ were spheres of radius
$r_{ij}/2$ in contact and spinning with angular velocities ${\bf
\omega}_i,{\bf \omega}_j$ the relative velocity at the ``surface'' of
the spheres would be equal to $\frac{1}{2}{\bf r}_{ij}\times({\bf
\omega}_i+{\bf \omega}_j)$. Then ${\bf F}^R$ gives a friction force
between the spheres proportional to this relative velocity. This force
produces an ``engaging'' effect in which neighbour particles prefer to
spin in opposite senses.

The last contribution in (\ref{sum3}) is a velocity-independent
stochastic force which is inspired by the tensorial structure of the
random forces that appear in the fluctuating hydrodynamics theory
\cite{lan59}. $\sigma$ is a parameter governing the overall noise
amplitude, the scalar functions
$\tilde{A}(r),\tilde{B}(r),\tilde{C}(r)$ define the range of the
random force, and we have introduced the following symmetric,
antisymmetric and traceless symmetric random matrices
\begin{eqnarray}
d{\bf W}^{S\mu\nu}_{ij}
&=&\frac{1}{2}\left[d{\bf W}^{\mu\nu}_{ij}
+d{\bf W}^{\nu\mu}_{ij}\right]
\nonumber \\
d{\bf W}^{A\mu\nu}_{ij}
&=&\frac{1}{2}\left[d{\bf W}^{\mu\nu}_{ij}
-d{\bf W}^{\nu\mu}_{ij}\right]
\nonumber \\
\overline{d{\bf W}}^S_{ij}&=&d{\bf W}^S_{ij}
-\frac{1}{D}{\rm tr}[d{\bf W}^S_{ij}]{\bf 1}
\label{decomp}
\end{eqnarray}
Here, $D$ is the physical dimension of space and $d{\bf
W}^{\mu\nu}_{ij}$ is a matrix of independent Wienner increments which
is assumed to be symmetric under particle interchange $d{\bf
W}^{\mu\nu}_{ij}=d{\bf W}^{\mu\nu}_{ji}$.  This symmetry will ensure
momentum conservation because $\tilde{\bf F}_{ij}=-\tilde{\bf
F}_{ji}$.  The matrix $d{\bf W}^{\mu\nu}_{ij}$ is an infinitesimal of
order $1/2$, and this is summarized in the Ito mnemotechnical rule
$d{\bf W}^{\mu\mu'}_{ii'}d{\bf W}^{\nu\nu'}_{jj'}=
\left[\delta_{ij}\delta_{i'j'}+\delta_{ij'}\delta_{ji'}\right]
\delta_{\mu\nu}\delta_{\mu'\nu'}dt$. From this stochastic property,
one derives straightforwardly the following rules from the different
parts

\begin{eqnarray}
{\rm tr}[d{\bf W}_{ii'}]{\rm tr}[d{\bf W}_{jj'}]
&=&\left[\delta_{ij}\delta_{i'j'}+\delta_{ij'}\delta_{ji'}\right]
Ddt
\nonumber\\
\overline{d{\bf W}}^{S\mu\mu'}_{ii'}
\overline{d{\bf W}}^{S\nu\nu'}_{jj'}
&=&\left[\delta_{ij}\delta_{i'j'}+\delta_{ij'}\delta_{ji'}\right]
\left[
\frac{1}{2}\left(\delta_{\mu\nu}\delta_{\mu'\nu'}+
\delta_{\mu\nu'}\delta_{\mu'\nu}\right)-\frac{1}{D}
\delta_{\mu\mu'}\delta_{\nu\nu'}
\right]dt
\nonumber\\
d{\bf W}^{A\mu\mu'}_{ii'}
d{\bf W}^{A\nu\nu'}_{jj'}&=&
\left[\delta_{ij}\delta_{i'j'}
+\delta_{ij'}\delta_{ji'}\right]
\frac{1}{2}\left(\delta_{\mu\nu}\delta_{\mu'\nu'}-
\delta_{\mu\nu'}\delta_{\mu'\nu}\right)dt
\nonumber\\
{\rm tr}[d{\bf W}_{ii'}]d{\bf W}^S_{jj'}
&=&{\rm tr}[d{\bf W}_{ii'}]d{\bf W}^A_{jj'}=
\overline{d{\bf W}}^{S\mu\mu'}_{ii'}
d{\bf W}^{A\nu\nu'}_{jj'}=0
\label{ran4}
\end{eqnarray}
These expressions show that the traceless symmetric, the trace and the
antisymmetric parts are independent stochastic processes. The
apparently complex structure of the random force is required in order
to be consistent with the tensor structure of the dissipative friction
forces.  This will become apparent when considering the associated
Fokker-Planck equation and requiring that it has a proper equilibrium
ensemble.

The force ${\bf F}_{ij}$ is the most general force that can be
constructed out of the vectors ${\bf r}_i,{\bf r}_j,{\bf v}_i,{\bf
v}_j,{\bf \omega}_i,{\bf \omega}_j$ and satisfies that : 1) it is
invariant under translational and Galilean transformations and
transforms as a vector under rotations; 2) it is linear in the linear
and angular velocities. This linearity is required in order to be
consistent with the Gaussian distribution of velocities at
equilibrium, as we will show later; 3) it satisfies Newton's third law
${\bf F}_{ij}=-{\bf F}_{ji}$ and, therefore, the total linear momentum
is a conserved quantity of the system.

\section{Fokker-Planck equation and equilibrium state}
The equations of motion (\ref{eqmotion}) are Langevin equations which
have associated a mathematically equivalent Fokker-Planck equation
\cite{gar83}.  The FPE governs the distribution function $\rho( r,
v,\omega; t)$ that gives the probability density that the $N$
particles of the system have specified values for the positions,
velocities and angular velocities.  Following standard \cite{gar83}
although quite tedious procedures, the FPE is given by

\begin{equation}
\partial_t \rho(r,v,\omega;t)
=\left[L^C+L^T+L^R\right]\rho(r,v,\omega;t)
\label{fp1}
\end{equation}
where

\begin{eqnarray}
L^C&=&-\left[\sum_i{\bf v}_i
\frac{\partial }{\partial {\bf r}_i}
+\sum_{i,j\neq i}\frac{1}{m}{\bf F}^C_{ij}
\frac{\partial}{\partial{\bf v}_i}\right]
\nonumber\\
L^T&=&
\sum_{i,j\neq i}\frac{\partial}{\partial {\bf v}_i}
\!\cdot\!\left[{\bf L}^T_{ij}+{\bf L}^R_{ij}\right]
\nonumber \\
L^R&=&-\frac{m}{I}
\sum_{i,j\neq i}\frac{\partial}{\partial {\bf \omega}_i}
\!\cdot\!\left(\frac{{\bf r}_{ij}}{2}\times
\left[{\bf L}^T_{ij}+{\bf L}^R_{ij}\right]\right)
\nonumber\\
{\bf L}^T_{ij}
&\equiv&-\frac{1}{m}{\bf F}^T_{ij}
+\frac{\sigma^2}{2}{\bf T}_{ij}\!\cdot\!
\left[\frac{\partial}{\partial{\bf v}_i}
-\frac{\partial}{\partial{\bf v}_j}\right]
\nonumber\\
{\bf L}^R_{ij}&\equiv&
-\frac{1}{m}{\bf F}^R_{ij}
+\frac{m}{I}\frac{\sigma^2}{2}{\bf T}_{ij}
\!\cdot\!
\left( \frac{{\bf r}_{ij}}{2}
\times \left[\frac{\partial}{\partial{\bf \omega}_i}
+\frac{\partial}{\partial{\bf \omega}_j}\right]\right)
\label{vecoper}
\end{eqnarray}
Here, the matrix ${\bf T}_{ij}$ is given by
\begin{equation}
{\bf T}_{ij}=
\frac{1}{2}\left[{\tilde A}^2(r_{ij})
+{\tilde C}^2(r_{ij})\right]{\bf 1}
+\left[\left(\frac{1}{2}-\frac{1}{D}\right)
{\tilde A}^2(r_{ij})+\frac{1}{D}{\tilde B}^2(r_{ij})
-\frac{1}{2}{\tilde C}^2(r_{ij})\right]
{\bf e}_{ij}{\bf e}_{ij}
\label{t}
\end{equation}

The steady state solution of equation (\ref{fp1}), $\partial_t
\rho=0$, gives the (unique) equilibrium distribution $\rho^{eq}$. We
now consider the conditions under which the steady state solution is
the Gibbs canonical ensemble
\begin{equation}
\rho^{eq}(r,v,\omega)=\frac{1}{Z}\exp\{-\left(\sum_i\frac{m}{2}v_i^2+
\frac{I}{2}\omega_i^2+V(r)\right)/k_BT\}
\label{eq}
\end{equation}
where $V$ is the potential function that gives rise to the
conservative forces ${\bf F}^C$, $k_B$ is Boltzmann's constant, $T$ is
the equilibrium temperature and $Z$ is the normalizing partition
function. We note that the velocity and angular velocity hydrodynamic
{\em fields} are Gaussian variables at equilibrium and, therefore, one
expects that the distribution function of the discrete values of
theses fields is also Gaussian.

The canonical ensemble is the equilibrium solution for the
conservative system, {\it i.e.}  $L^C\rho^{eq}=0$. If, in addition,
the following equations are satisfied ${\bf L}^T_{ij}\rho^{eq}={\bf
L}^R_{ij}\rho^{eq}=0$ then we will have $L\rho^{eq}=0$ and the Gibbs
equilibrium ensemble will be the unique stationary solution of the
dynamics.  These equations will be satisfied if the detailed balance
condition $\gamma =\frac{\sigma^2m}{2k_BT}$ is satisfied and also
${\bf M}^R({\bf r}_{ij})={\bf M}^T({\bf r}_{ij})={\bf T}_{ij}$. This
implies

\begin{eqnarray}
A(r)&=&\frac{1}{2}\left[{\tilde A}^2(r)+{\tilde C}^2(r)\right]
\nonumber\\
B(r)&=&
\frac{1}{2}\left[{\tilde A}^2(r)-{\tilde C}^2(r)\right]
+\frac{1}{D}\left[{\tilde B}^2(r)-{\tilde A}^2(r)\right]
\label{AB}
\end{eqnarray}
We observe, therefore, that the initial hypothesis for the tensorial
structure of the dissipative and random forces was correct and
consistent with equilibrium statistical mechanics.

The structure of the dissipative forces postulated in the fluid
particle model (disregarding angular variables) is essentially the
same as the structure of the viscous forces in the (isothermal) SPD
algorithm. By using the discretization of Takeda {\em et al.}
\cite{tak94} the correspondence is

\begin{eqnarray}
V(r)&=&2\frac{p_0}{mn_0^2}W(r)
\nonumber\\
\gamma m A(r)&=&\frac{1}{mn_0^2}
\left[\eta W''(r)
+\left[2\eta+\left(\zeta+\frac{\eta}{3}\right)\right]
\frac{W'(r)}{r}\right]
\nonumber\\
\gamma m B(r)&=&\frac{1}{mn_0^2}
\left(\zeta+\frac{\eta}{3}\right)\left[ W''(r)-
\frac{W'(r)}{r}\right]
\label{corres}
\end{eqnarray}
where $p_0,n_0$ are the equilibrium pressure and number
density, and $W(r)$ is the bell-shaped weight function used in the
discretization of the Navier-Stokes equation (the assumption that the
density of all particles is almost constant has been taken).  The SPD
algorithm does not conserve angular momentum and does not incorporate
thermal fluctuations. The first issue can be reduced at the cost of
increasing the resolution (i.e. by increasing the computational cost
of the simulations \cite{tak94}). Regarding the second issue, we have
formulated in this letter how to introduce consistently the thermal
noise ({\em i.e.} by selecting
$\tilde{A}(r),\tilde{B}(r),\tilde{C}(r)$ satisfying (\ref{AB})).
However, we note a serious problem in the expression for the scalar
function $A(r)$ in terms of the weight function $W(r)$ in
(\ref{corres}).  The left hand side of the second equation in
(\ref{corres}) is negative for some values of $r$ for a bell-shaped
weight function $W(r)$. This is unacceptable in view of
Eqn. (\ref{AB}). The smoothed particle dynamics algorithm does not
allow, then, for a consistent introduction of thermal noise, at least
in the form presented in Ref. \cite{tak94}.

In summary, we have proposed a fluid particle model by introducing
shear forces and spin into the original DPD algorithm. The model has
the correct equilibrium state and has a structure similar to SPD, but
with angular momentum conservation and correct thermal
fluctuations. In this way, the connection between DPD and SPD is
clarified.

This work has been partially supported by a DGICYT Project No
PB94-0382 and by E.C. contract ERB-CHRXCT-940546.

\end{document}